\documentclass[conference]{IEEEtran}

\ifCLASSINFOpdf
\else
\fi
\usepackage{booktabs}
\usepackage{amsmath,amssymb,amsfonts}
\usepackage{algorithmic}
\usepackage{graphicx}
\usepackage{textcomp}
\usepackage{xcolor}
\usepackage{hyperref}
\usepackage{array}
\usepackage{siunitx}
\usepackage{booktabs}
\usepackage{float}
\usepackage{hyperref}
\usepackage{listings} 
\usepackage{color} 
\usepackage{xcolor}
\usepackage{hyperref}
\definecolor{codegreen}{rgb}{0,0.6,0}
\definecolor{codegray}{rgb}{0.5,0.5,0.5}
\definecolor{codepurple}{rgb}{0.58,0,0.82}
\definecolor{backcolour}{rgb}{0.95,0.95,0.92}
\definecolor{codeblue}{rgb}{0.25, 0.5, 0.75}

\usepackage{listings}
\usepackage{xcolor}
\usepackage{array}

\lstdefinestyle{mystyle2}{
    backgroundcolor=\color{backcolour},   
    commentstyle=\color{green},
    keywordstyle=\color{blue},
    stringstyle=\color{purple},
    basicstyle=\ttfamily\footnotesize, 
    breakatwhitespace=false,         
    breaklines=true,                 
    captionpos=b,                    
    keepspaces=true,                 
    numbers=none,                                  
    showspaces=false,                
    showstringspaces=false,
    showtabs=false,                  
    tabsize=1
}

\lstdefinestyle{mystyle}{
    backgroundcolor=\color{backcolour},   
    commentstyle=\color{codeblue},
    keywordstyle=\color{magenta},
    stringstyle=\color{codepurple},
    basicstyle=\ttfamily\footnotesize,
    breakatwhitespace=false,         
    breaklines=true,                 
    captionpos=b,                    
    keepspaces=true,                 
    numbers= none,                           
    showspaces=false,                
    showstringspaces=false,
    showtabs=false,                  
    tabsize=1
}

\begin{document}

\title{Unravelling Technical debt topics through Time, Programming Languages and Repository}


\author{
    \IEEEauthorblockN{Karthik Shivashankar}
    \IEEEauthorblockA{University of Oslo\\
    karths@ifi.uio.no}
    \and
        \IEEEauthorblockN{Antonio Martini}
    \IEEEauthorblockA{University of Oslo\\
    antonima@ifi.uio.no}
}

\maketitle

\begin{abstract}
This study explores the dynamic landscape of Technical Debt (TD) topics in software engineering by examining its evolution across time, programming languages, and repositories. Despite the extensive research on identifying and quantifying TD, there remains a significant gap in understanding the diversity of TD topics and their temporal development. To address this, we have conducted an explorative analysis of TD data extracted from GitHub issues spanning from 2015 to September 2023. We employed BERTopic for sophisticated topic modelling. This study categorises the TD topics and tracks their progression over time. Furthermore, we have incorporated sentiment analysis for each identified topic, providing a deeper insight into the perceptions and attitudes associated with these topics. This offers a more nuanced understanding of the trends and shifts in TD topics through time, programming language, and repository.
\end{abstract}

\IEEEpeerreviewmaketitle

\section{Introduction}
Technical Debt (TD) in software engineering is a metaphorical concept that reflects the extra developmental work arising when code that is easy to implement in the short run is used instead of applying the best overall solution. This phenomenon, first coined by Cunningham in 1992, has become a central topic in software engineering research and practice \cite{Cunningham1992}. Despite extensive research on identifying and quantifying TD, there remains a significant gap in understanding the diversity and temporal development of TD topics. This gap is particularly evident in the dynamic and evolving nature of software development practices, influenced by factors such as technological advancements, programming language trends, and the changing nature of software projects.
To address this gap, our study conducts an explorative analysis of TD data extracted from GitHub issues spanning from 2015 to September 2023. GitHub, a widely used platform for software development, offers a rich repository of real-world data on software development practices, including discussions and resolutions of TD issues \cite{Kalliamvakou2014}. Using BERTopic, we explore sophisticated topic modelling to categorise and track the progression of TD topics over time \cite{Grootendorst2020}. Additionally, our study incorporates sentiment analysis for each identified topic, providing deeper insights into the perceptions and attitudes associated with these topics \cite{Liu2012}.
With our curated dataset, we can also identify TD issue topics in relation to the programming languages used in the repositories. This approach allows for the identification of language-specific TD patterns. For instance, specific TD issues may be more prevalent in dynamically typed languages like Python, while others might be more common in statically typed languages like Java. This aligns with the findings of Spinellis et. al. \cite{Spinellis2020}, who discussed the impact of programming language characteristics on software quality and maintenance. Understanding the nuances of Topics discovered in TD in relation to programming languages can significantly enhance the strategies for managing TD. 
The motivation for this study stems from the critical need to understand the evolving landscape of TD in software engineering. As software systems grow in complexity and scale, the management of TD becomes increasingly challenging and vital for the sustainability and success of software projects \cite{TOM20131498}. However, existing literature primarily focuses on identifying and quantifying TD, with less emphasis on the diversity and evolution of TD topics over time. This gap is significant in the context of rapidly changing software development paradigms influenced by new technologies, programming languages, and methodologies.
Furthermore, the emotional and perceptual aspects of TD are often overlooked in technical analyses. Developers' sentiments and attitudes towards various TD issues can significantly impact decision-making processes and prioritisation in software maintenance and evolution \cite{10.1145/2491411.2494578}. Therefore, a comprehensive analysis that combines topic modelling with sentiment analysis is essential to gain a holistic understanding of TD in the software development lifecycle. Our research is primarily driven by these Research Questions (RQ):
\begin{itemize}
    \item RQ1: What variety of topics related to TD are commonly identified in software project issues discussion?
    \item RQ2: How have the identified topics of Technical Debt evolved over time in the context of different programming languages and repositories?
    \item RQ3: In what ways do the topics and prevalent sentiments in TD vary across different programming languages and repositories?
\end{itemize}
These questions seek to uncover the range of TD issues that software developers encounter and track their progression or changes over the years.
\paragraph{Contribution}
This study offers significant contributions to understanding Technical Debt (TD) in software engineering. It utilizes advanced topic modeling with BERTopic to identify a diverse range of TD topics from GitHub issues, providing a more nuanced categorization than traditional methods. The research also examines the relationship between TD issues and programming languages, shedding light on how TD varies across different languages. Additionally, it presents a temporal analysis of TD topics from 2015 to September 2023, highlighting the evolution of TD over time. Finally, the study integrates sentiment analysis, offering insights into the emotional context and perceptions of developers regarding TD topics.
 \section{Background}
\subsection{Topic Modelling and BERTopic}
Topic modelling is a technique in natural language processing and machine learning that uncovers latent topics in document collections. It helps summarize and understand large volumes of textual data by identifying underlying themes or topics. Traditional methods like Latent Dirichlet Allocation (LDA) and Latent Semantic Analysis (LSA) are commonly used, relying on statistical models to deduce the hidden thematic structure of document collections.
BERTopic marks a significant advancement in topic modelling. It enhances traditional methods by using transformer-based language models, particularly BERT (Bidirectional Encoder Representations from Transformers) \cite{Devlin2018}, to create document embeddings. These embeddings more effectively capture the semantic meaning of texts.
BERTopic is renowned for generating more coherent and interpretable topics than classical models like LDA \cite{Grootendorst2020}. It's particularly effective in handling short texts and texts from various domains, offering flexibility and domain adaptation \cite{deGroot2022GeneralizabilityBERTopic}. Unlike other topic modelling techniques like LSA, NMF, and PAM, BERTopic consistently yields more meaningful topics in various applications, including analyzing customer reviews \cite{Krishnan2023TopicModelingCustomerReviews}.
Furthermore, the effectiveness of the transformer-based model in supervised learning has already been well established by Skryseth et al. \cite{10207085}  in automatically classifying TD issues from GitHub and by  Li et al. \cite{Li2023}, who utilized BERT for estimating the effort required to repay Self-Admitted Technical Debt (SATD) in software projects. 
\section{Methodology}
The methodology of this study is structured into three main components: Data Collection, BERTopic for Topic Modelling, and Sentiment Analysis. Each component plays a crucial role in achieving the study's objectives.
\subsection{Data Collection}
 The methodology for data extraction involves identifying GitHub issues explicitly tagged with the keyword '*debt*' labels by developers. This targeted approach ensures the relevance of the data to TD. The extraction process was designed to avoid duplicates, ensuring the integrity and quality of the dataset. Using this approach, we have mined around 45,000 TD-related issues. This data collection method aligns with the practices recommended by  Skryseth et al. \cite{10207085}. The historical span of the data extracted is from 2015 to September 2023, providing a comprehensive view of the trends and developments in TD issues within software development. This extensive temporal coverage is crucial for analysing the evolution of TD, offering insights into the shifting paradigms and practices in software development over the years. This approach aligns with the findings of  \cite{Kalliamvakou2014}, who highlight the significance of GitHub as a platform for software development research.

A vital component of this research was using the GitHub Archive \cite{GitHubArchive}, an extensive resource archiving public GitHub activities. It enabled the collection of a broader range of TD discussions beyond the explicit use of the word '*debt.*' by using regular expressions from the issues event, encompassing phrases like 'technical debt', 'code debt', and 'debt reduction'. This enhanced the dataset with a more comprehensive array of TD instances over the selected timeframe.

Moreover, the GitHub API \cite{GitHubRESTAPI2022} was utilised to extract data on the primary programming languages in repositories linked to the Self-Admitted Technical Debt (SATD) issues. This added dimension provided critical context for analyzing TD in relation to different programming environments. Understanding language-specific TD patterns and practices is essential for a holistic analysis, as it allows for examining the occurrence, discussion, and technical context of TD.
Overall, this data collection offers a thorough and multifaceted approach to understanding TD in software development, combining historical data mining, linguistic diversity in TD identification, and programming language analysis to provide comprehensive insights into the evolution and management of Technical Debt.

\subsection{Topic Modelling}
BERTopic is employed for topic modelling in this study. BERT's ability to understand the context of words in the text by considering the words that come before and after the sentence makes it highly effective for unsupervised learning. BERTopic uses this capability to group similar textual content, enabling the identification of prevalent themes in TD discussions. This process begins with document embedding, where documents are transformed into embeddings using pre-trained transformer-based language models capturing essential contextual relationships within the text. This is crucial for understanding the nuances of TD-related discussions. The next step is clustering; these embeddings are grouped to identify distinct themes in TD, aiding in revealing major trends and concerns in the field. Finally, BERTopic employs a class-based variation of TF-IDF (Term Frequency-Inverse Document Frequency) for coherent topic representation, identifying the most representative terms for each topic \cite{Grootendorst2020}.
\subsection{Sentiment Analysis}
In our study, sentiment analysis is integrated with topic modelling to understand the emotional context and attitudes towards various aspects of TD. This analysis, crucial for comprehending the nuances of TD, is based on methodologies that classify text into positive, negative, or neutral sentiments, as described by Liu et al. \cite{Liu2012}. This approach aligns with the research by Silva et al. \cite{Silva2016}, who emphasized the significance of emotional and perceptual aspects of TD in the context of pull request acceptance in software projects. Our sentiment analysis aims to capture the emotional context surrounding TD topics, offering insights into developers' perceptions and attitudes.
Furthermore, the study employs the VADER (Valence Aware Dictionary and sEntiment Reasoner) sentiment analysis tool, a lexicon and rule-based framework adept at analyzing sentiments in social media contexts. VADER's effectiveness in both general and domain-specific sentiment analysis tasks is well-documented in various studies \cite{HuttoGilbert2014, Zhang2019, Li2021, Dong2014, MutluOzgur2022, KaushikMishra2014}. By integrating VADER sentiment scores with topic modelling for each GitHub issue, we gain a deeper understanding of the sentiment dynamics within different TD topics, enhancing our comprehension of how these issues are perceived and discussed in software development projects.
VADER (Valence Aware Dictionary and sEntiment Reasoner) is a sentiment analysis tool that categorizes text into positive, negative, and neutral sentiments, along with a composite score. Positive, negative, and neutral scores range from 0 to 1, indicating the proportion of text in each sentiment category. The compound score, normalized between -1 and +1, sums the valence scores of each word, providing a single measure of overall sentiment. Typically, thresholds for the compound score classify text as positive, negative, or neutral. In analyzing Technical Debt discussions on GitHub, these scores offer insights into the emotional tone, with the compound score simplifying overall sentiment classification.

\subsection{Replication package}

A comprehensive replication package is provided to facilitate reproducibility and support further research in the field of Technical Debt (TD). This package includes all relevant data, code, and graphical representations pertaining to the detailed analysis of TD, which is available in Zenodo \href{https://zenodo.org/records/10148205}{Replication package link} \footnote{https://zenodo.org/records/10148205}. It encompasses topic modelling, sentiment analysis, and temporal analysis, reflecting the evolution of TD topics and sentiments from 2015 to September 2023. The dataset within this package is meticulously segmented by programming languages, such as Python, TypeScript, JavaScript, Java, C++, Rust, Go, CSharp, and HTML, and includes specific repositories like 'microsoft/vscode', 'metabase/metabase', and 'elastic/kibana'. This segmentation facilitates the exploration of TD topics unique to each language and repository. In addition to that, the inclusion of sentiment analysis results provides insights into developers' attitudes towards TD issues.

\section{Results}
\subsection{RQ1: Identified Topics of Technical Debt}
 The application of BERTopic revealed various topics within TD discussions. These topics included issues related to different topics like testing, UI components, and more, as listed in Table \ref{tab:td_issues}, as well as the count of  TD issues belonging to these identified topics.  The various topics identified underscores the multifaceted nature of Technical Debt in SE. In unsupervised learning, the algorithm autonomously discovers patterns and relationships in the data without any predefined labels or categories. In the context of BERTopic, the tool analyzes text data related to Technical Debt and independently identifies distinct topics based on the patterns of term usage within the dataset.
 The topics listed in Table \ref{tab:td_issues}, generated through BERTopic using TF-IDF (Term Frequency-Inverse Document Frequency), represent distinct areas of focus within Technical Debt (TD) discussions in Software Engineering. BERTopic, by employing TF-IDF, identifies the most representative and relevant terms for each topic based on their frequency and distinctiveness within the dataset. Here's a breakdown of how these topics and their names are determined. For instance, let's take the first two topics in Table \ref{tab:td_issues}
Testing (3408 TD issues): The topic name "testing\_tests\_test\_tested" is formed because these terms frequently appear in discussions related to testing practices, methodologies, and issues. The high count indicates a significant focus on testing-related TD.
UI Components (1810 issues): "component\_ui\_components\_css" suggests a focus on user interface components and styling (CSS), reflecting the technical considerations and debt associated with UI development.

\begin{table}[h] 
\centering 
\caption{Top 5 Topics identified} 
\label{tab:td_issues} 
\begin{tabular}{rrl}
\toprule
Topic & Count TD Issues & Topic Name Assigned \\
1 & 3408 & testing\_tests\_test\_tested \\
2 & 1810 & component\_ui\_components\_css \\
3 & 1544 & refactor\_remove\_redux\_refactoring \\
4 & 1204 & fsr\_issue\_qa \\
5 & 544 & aws\_resource\_acctest\_issue \\

\bottomrule
\end{tabular}
\end{table}

\subsection{RQ2: Evolution of TD Topics Over Time}
\begin{figure*}
    \centering
    \includegraphics[width=1\linewidth]{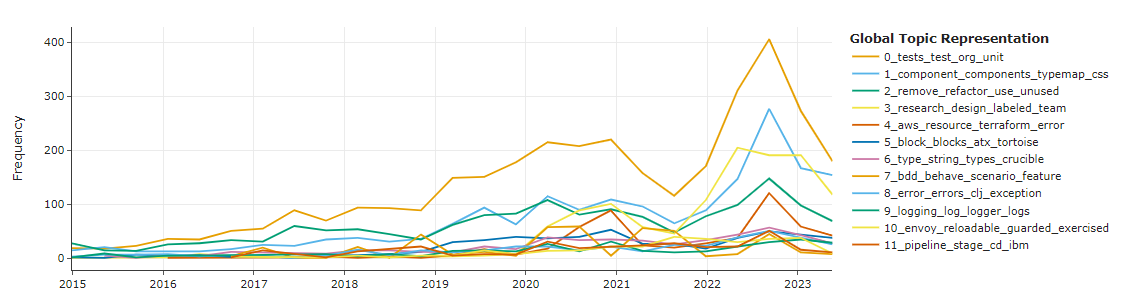}
    \caption{Main Topics from the entire dataset and its evolution with respect to time }
    \label{fig:1}
\end{figure*}

\begin{figure*}
    \centering
    \includegraphics[width=1\linewidth]{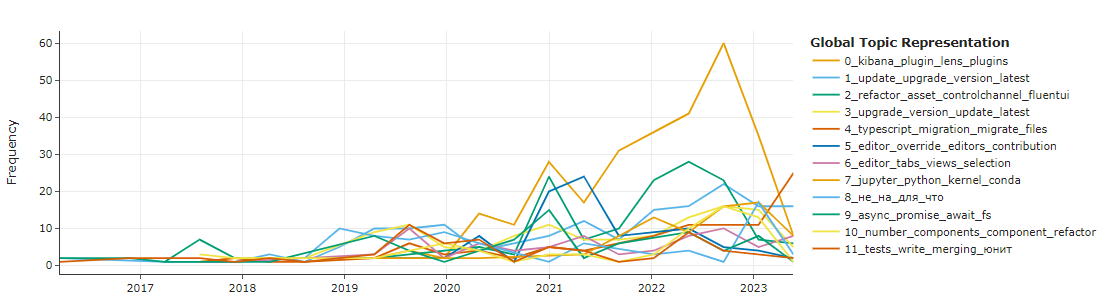}
    \caption{Main Topics from the filtered TypeScript programming language   and its evolution with respect to time }
    \label{fig:2}
\end{figure*}

\begin{figure*}
    \centering
    \includegraphics[width=1\linewidth]{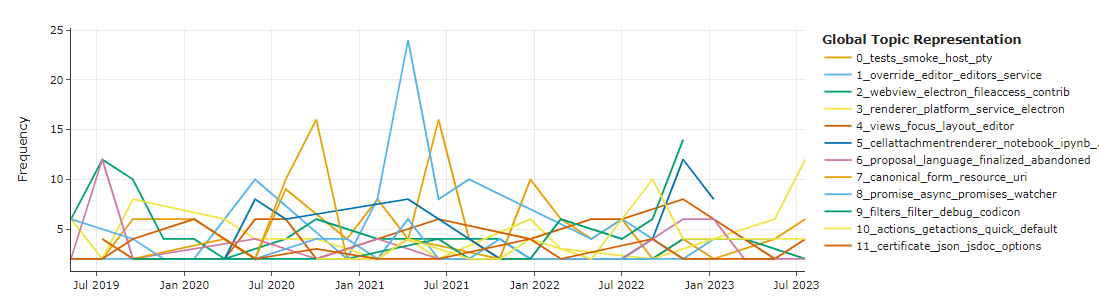}
    \caption{Main Topics from the filtered VSCode repository  and its evolution with respect to time }
    \label{fig:3}
\end{figure*}

 Our analysis showed that specific topics, such as Tests and  UI components, refactoring, CI/CD  and cloud-native-related TD issues from the entire dataset, have become more prominent see Figure \ref{fig:1}. This shift could reflect modern software development's evolving complexities and scaling challenges as noted by Bogner et al \cite{Bogner2019}. \\
Similarly, Figure \ref{fig:2} shows the evolution of TD-related topics concerning the TypeScript programming language. Here, we can infer that many of the TD-related topics identified are related to updating software packages,  async communication, UI components, and testing, which reflect the Typescript landscape and its usage in modern web development.\\
Figure \ref{fig:3} shows the evolution of TD related topics concerning the VSCode editor, here we can infer that in VScode, most of the TD issues are related to Testing, Webview, Electron and other related Technologies or frameworks on which VS code is built, apart from that we can also see language specific TD issues like Promise, Async  and JSON,  given Vscode is mostly build on TypeScript. \\
\subsection{RQ3: Sentiment Analysis Results}
\begin{figure}
    \centering
    \includegraphics[width=0.75\linewidth]{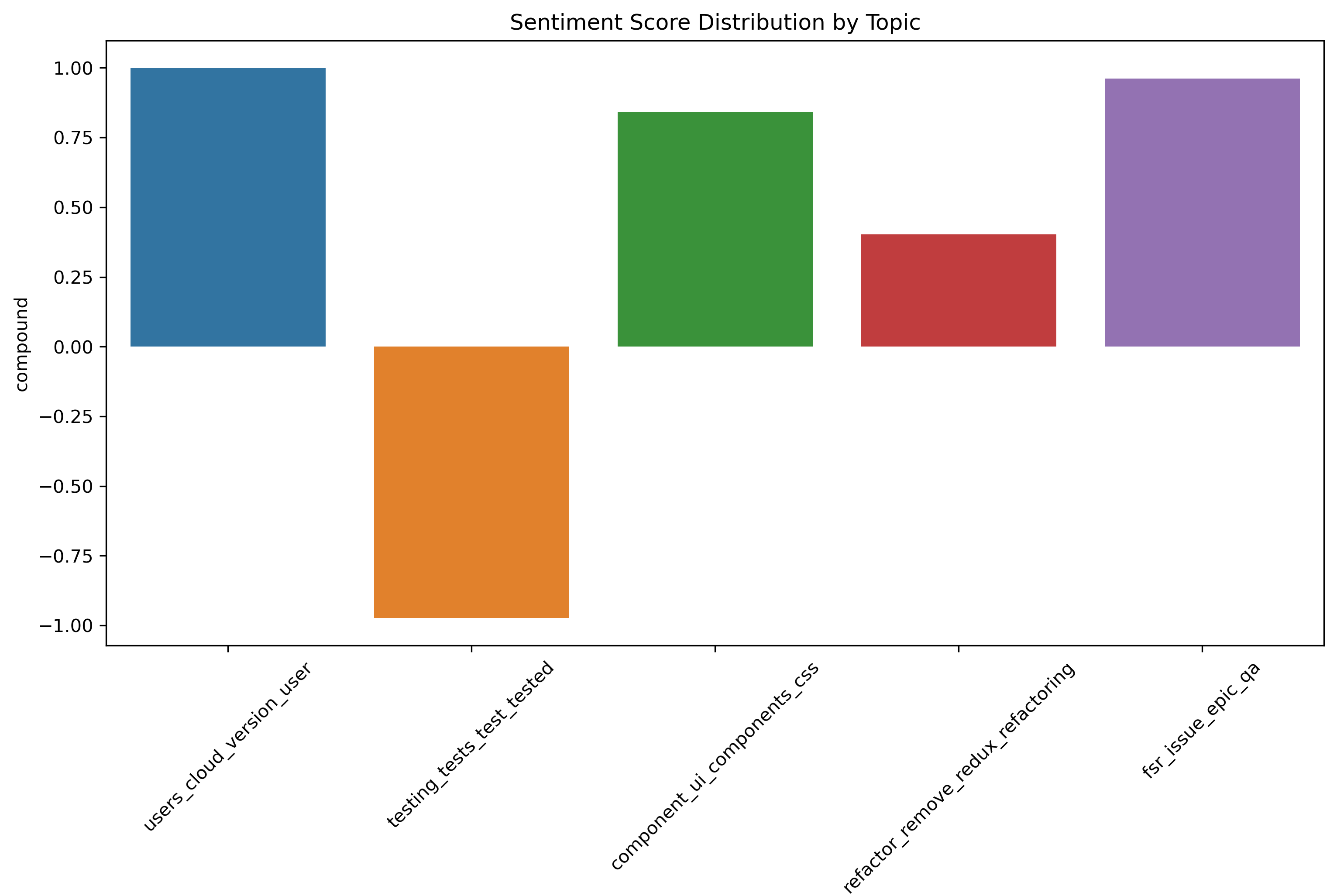}
    \caption{VADER Compound Sentiment Score for the Entire Dataset with respect to each top Topics}
    \label{fig:4}
\end{figure}

\begin{figure}
    \centering
    \includegraphics[width=0.75\linewidth]{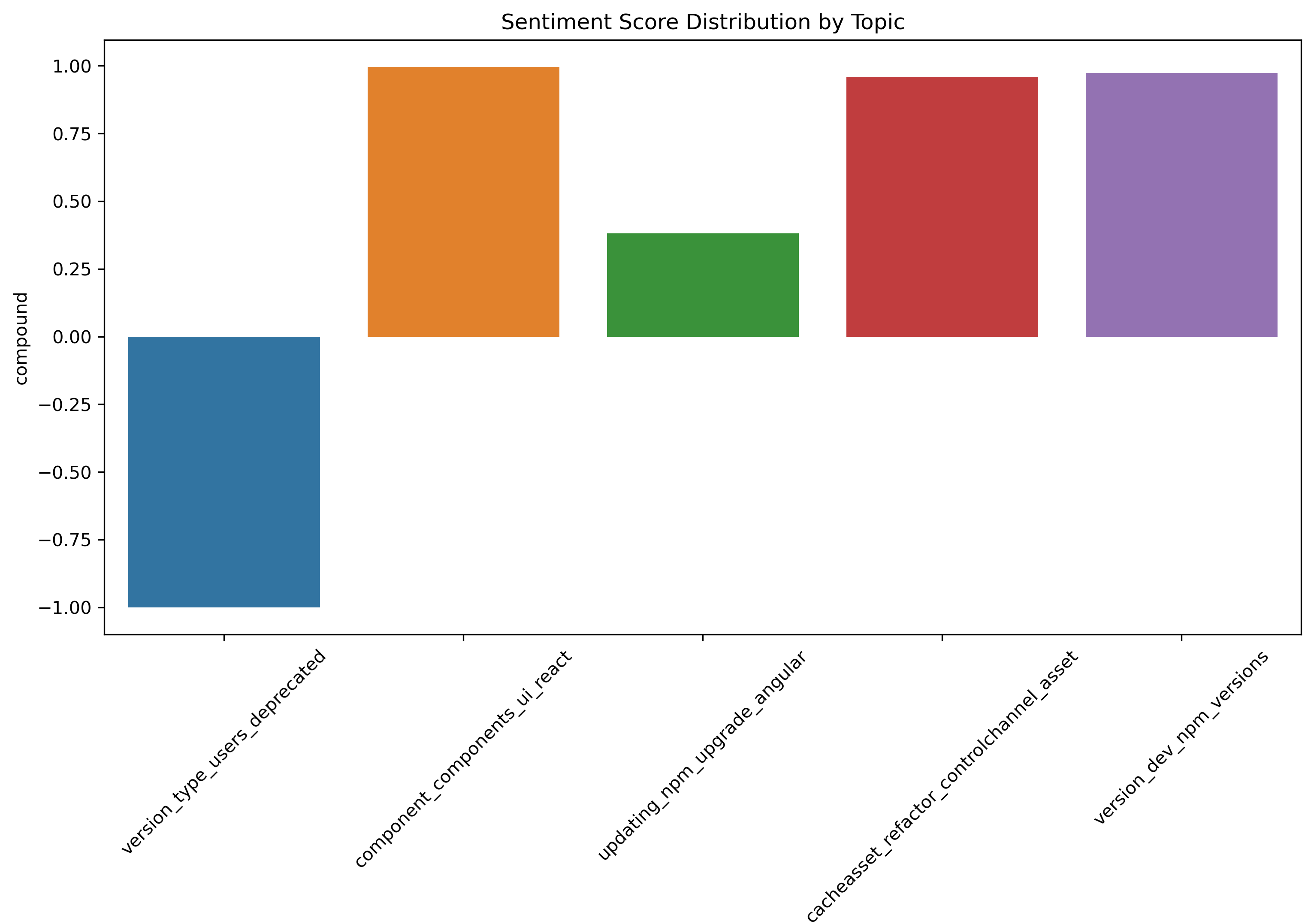}
    \caption{VADER Compound Sentiment Score for the filtered TypeScript  dataset with respect to each top Topics}
    \label{fig:5}
\end{figure}

 The sentiment analysis provided insights into how developers perceive various TD topics. For example, issues related to legacy code integration and inadequate testing were often associated with negative sentiments, reflecting frustration and concern. This finding is consistent with the research by Huang et al.  \cite{Huang2021}, who highlighted the impact of sentiments on software development practices. The Compound score in VADER (Valence Aware Dictionary and sEntiment Reasoner) is a metric used to gauge the overall sentiment of a text. A score of -1 indicates extremely negative sentiment, while +1 indicates extremely positive sentiment. A neutral sentiment score near 0 typically indicates neutral sentiment. 
Figure \ref{fig:4} shows the Compound Sentiment score for the main TD topic identified for the entire dataset. Here, we can infer that Testing topics are more inclined towards negative sentiments.  The tone or sentiments for other identified TD topics are more inclined towards  positive sentiment  with varying degrees with some neutral sentiment in Refactoring. \\
Similarly, Figure \ref{fig:5} shows the Compound Sentiment score for the main TD topic identified for the  TypeScript programming Language. Here, we can infer that package deprecation and version-related issues likely related to the dependency and package management  topics are more inclined towards negative sentiments.  The tone or sentiments for other identified TD topics are more inclined towards  positive sentiment with varying degrees with some neutral sentiment in the Updating and Upgrading  NPM package.  \\
\section{Discussion:}
The study reveals Topics identified from the  TD issues, highlighting key areas where TD  is more prevalent. It tracks the evolution of these topics over time, focusing on various programming languages (Python, TypeScript, JavaScript, Java, C++, Rust, Go, C\#, HTML) and specific repositories ('microsoft/vscode', 'metabase/metabase', 'elastic/kibana'); see replication packages for all the data and code. This approach uncovers prevalent issues in TD and how they adapt to technological changes and development practices.

The sentiment analysis provides insights into developers' emotional responses to TD issues. It shows a range of sentiments, with negative feelings prominent in areas like testing and package management, reflecting the challenges developers face. In contrast, more neutral or positive sentiments in other areas suggest greater confidence in managing TD. These findings are crucial for understanding the impact of TD on developers' psychological well-being and can guide strategies to enhance their experience and productivity.

However, the study acknowledges several limitations. The use of BERTopic, while effective, may introduce subjective biases based on the issue description for TD in topic identification. Based on a specific set of repositories containing SATD issues, the analysis might not represent the broader spectrum of TD issues. The evolution analysis relies on existing data from TD issues mined from the GitHub Archive, thereby possibly overlooking recent trends. VADER's sentiment analysis, although useful, may not fully capture the complexity of emotional expressions in technical discussions. Additionally, there's a potential risk of interpretation bias in analyzing topics and sentiments, which could affect the conclusions.

\section{Conclusion}
This study offers a novel insight into the landscape of Technical Debt in software engineering, employing BERTopic for an in-depth analysis of TD topics. The findings reveal a diverse and evolving range of issues encompassed under Technical Debt, highlighting the significance of understanding these topics for effective software development and maintenance. By integrating sentiment analysis, the study not only identifies what TD topics are discussed but also captures the emotional tone surrounding these discussions, this study not only contributes to the existing body of knowledge but also opens avenues for further exploration and understanding in this vital area of TD related topics in software development.

\bibliographystyle{IEEEtran}
\bibliography{sample-base}

\begin{thebibliography}{10}
\providecommand{\url}[1]{#1}
\csname url@samestyle\endcsname
\providecommand{\newblock}{\relax}
\providecommand{\bibinfo}[2]{#2}
\providecommand{\BIBentrySTDinterwordspacing}{\spaceskip=0pt\relax}
\providecommand{\BIBentryALTinterwordstretchfactor}{4}
\providecommand{\BIBentryALTinterwordspacing}{\spaceskip=\fontdimen2\font plus
\BIBentryALTinterwordstretchfactor\fontdimen3\font minus \fontdimen4\font\relax}
\providecommand{\BIBforeignlanguage}[2]{{%
\expandafter\ifx\csname l@#1\endcsname\relax
\typeout{** WARNING: IEEEtran.bst: No hyphenation pattern has been}%
\typeout{** loaded for the language `#1'. Using the pattern for}%
\typeout{** the default language instead.}%
\else
\language=\csname l@#1\endcsname
\fi
#2}}
\providecommand{\BIBdecl}{\relax}
\BIBdecl

\bibitem{Cunningham1992}
W.~Cunningham, ``The wycash portfolio management system,'' in \emph{OOPSLA '92 Addendum to the Proceedings}, 1992.

\bibitem{Kalliamvakou2014}
E.~Kalliamvakou, G.~Gousios, K.~Blincoe, L.~Singer, D.~M. German, and D.~Damian, ``The promises and perils of mining github,'' in \emph{Proceedings of the 11th Working Conference on Mining Software Repositories}.\hskip 1em plus 0.5em minus 0.4em\relax ACM, 2014, pp. 92--101.

\bibitem{Grootendorst2020}
M.~Grootendorst, ``Bertopic: Leveraging bert and c-tf-idf to create easily interpretable topics,'' \emph{arXiv preprint arXiv:2012.14210}, 2020.

\bibitem{Liu2012}
B.~Liu, \emph{Sentiment Analysis and Opinion Mining}.\hskip 1em plus 0.5em minus 0.4em\relax Synthesis Lectures on Human Language Technologies, 2012, vol.~5, no.~1.

\bibitem{Spinellis2020}
D.~Spinellis, ``The effect of programming languages on software quality and maintenance: A literature review,'' \emph{arXiv preprint arXiv:2007.07568}, 2020.

\bibitem{TOM20131498}
\BIBentryALTinterwordspacing
E.~Tom, A.~Aurum, and R.~Vidgen, ``An exploration of technical debt,'' \emph{Journal of Systems and Software}, vol.~86, no.~6, pp. 1498--1516, 2013. [Online]. Available: \url{https://www.sciencedirect.com/science/article/pii/S0164121213000022}
\BIBentrySTDinterwordspacing

\bibitem{10.1145/2491411.2494578}
\BIBentryALTinterwordspacing
E.~Guzman and B.~Bruegge, ``Towards emotional awareness in software development teams,'' in \emph{Proceedings of the 2013 9th Joint Meeting on Foundations of Software Engineering}, ser. ESEC/FSE 2013.\hskip 1em plus 0.5em minus 0.4em\relax New York, NY, USA: Association for Computing Machinery, 2013, p. 671–674. [Online]. Available: \url{https://doi.org/10.1145/2491411.2494578}
\BIBentrySTDinterwordspacing

\bibitem{Devlin2018}
J.~Devlin, M.-W. Chang, K.~Lee, and K.~Toutanova, ``Bert: Pre-training of deep bidirectional transformers for language understanding,'' \emph{arXiv preprint arXiv:1810.04805}, 2018.

\bibitem{deGroot2022GeneralizabilityBERTopic}
\BIBentryALTinterwordspacing
M.~de~Groot, M.~Aliannejadi, and M.~R. Haas, ``Experiments on generalizability of bertopic on multi-domain short text,'' 2022. [Online]. Available: \url{http://arxiv.org/abs/2212.08459v1}
\BIBentrySTDinterwordspacing

\bibitem{Krishnan2023TopicModelingCustomerReviews}
\BIBentryALTinterwordspacing
A.~Krishnan, ``Exploring the power of topic modeling techniques in analyzing customer reviews: A comparative analysis,'' 2023. [Online]. Available: \url{http://arxiv.org/abs/2308.11520v1}
\BIBentrySTDinterwordspacing

\bibitem{10207085}
D.~Skryseth, K.~Shivashankar, I.~Pilán, and A.~Martini, ``Technical debt classification in issue trackers using natural language processing based on transformers,'' in \emph{2023 ACM/IEEE International Conference on Technical Debt (TechDebt)}, 2023, pp. 92--101.

\bibitem{Li2023}
Y.~Li, M.~Soliman, and P.~Avgeriou, ``Automatically estimating the effort required to repay self-admitted technical debt,'' \emph{arXiv preprint arXiv:2309.06020}, 2023.

\bibitem{GitHubArchive}
``Github archive,'' \url{https://www.gharchive.org/}, 2023, accessed: 2023-11-17.

\bibitem{GitHubRESTAPI2022}
``Github rest api documentation,'' \url{https://docs.github.com/en/rest}, 2022, accessed: 2022-11-28.

\bibitem{Silva2016}
M.~C.~O. Silva, M.~T. Valente, and R.~Terra, ``Does technical debt lead to the rejection of pull requests?'' \emph{arXiv preprint arXiv:1604.01450}, 2016.

\bibitem{HuttoGilbert2014}
C.~Hutto and E.~Gilbert, ``Vader: A parsimonious rule-based model for sentiment analysis of social media text,'' \emph{Eighth International AAAI Conference on Weblogs and Social Media}, 2014.

\bibitem{Zhang2019}
Y.~Zhang \emph{et~al.}, ``Scenariosa: A large scale conversational database for interactive sentiment analysis,'' \emph{arXiv preprint arXiv:1907.05562}, 2019.

\bibitem{Li2021}
C.~Li \emph{et~al.}, ``Sentiprompt: Sentiment knowledge enhanced prompt-tuning for aspect-based sentiment analysis,'' \emph{arXiv preprint arXiv:2109.08306}, 2021.

\bibitem{Dong2014}
L.~Dong \emph{et~al.}, ``A statistical parsing framework for sentiment classification,'' \emph{arXiv preprint arXiv:1401.6330}, 2014.

\bibitem{MutluOzgur2022}
M.~M. Mutlu and A.~Özgür, ``A dataset and bert-based models for targeted sentiment analysis on turkish texts,'' \emph{arXiv preprint arXiv:2205.04185}, 2022.

\bibitem{KaushikMishra2014}
C.~Kaushik and A.~Mishra, ``A scalable, lexicon based technique for sentiment analysis,'' \emph{arXiv preprint arXiv:1410.2265}, 2014.

\bibitem{Bogner2019}
J.~Bogner, J.~Fritzsch, S.~Wagner, and A.~Zimmermann, ``Assuring the evolvability of microservices: Insights into industry practices and challenges,'' \emph{arXiv preprint arXiv:1906.05013}, 2019.

\bibitem{Huang2021}
Z.~Huang, Z.~Shao, G.~Fan, J.~Gao, Z.~Zhou, K.~Yang, and X.~Yang, ``Predicting community smells' occurrence on individual developers by sentiments,'' \emph{arXiv preprint arXiv:2103.07090}, 2021.

\end{thebibliography}

\end{document}